\documentclass[useAMS,usenatbib]{mnras}
\usepackage{times}
\usepackage{graphicx}
\usepackage{epsfig}
\usepackage[T1]{fontenc}
\usepackage{adjustbox}
\usepackage{ifpdf}
\usepackage[varg]{txfonts}
\usepackage{hyperref}
\usepackage{deluxetable}

\newcommand{\src}{IGR~J20084}

\def\amin{\ifmmode^{\prime}\else$^{\prime}$\fi}
\def\asec{\ifmmode^{\prime\prime}\else$^{\prime\prime}$\fi}

\def\simgt{\lower.5ex\hbox{$\; \buildrel > \over \sim \;$}}
\def\simlt{\lower.5ex\hbox{$\; \buildrel < \over \sim \;$}}

\newcommand\chandra{{\it Chandra}}
\newcommand\xmm{{\it XMM-Newton}}

\newcommand\integral{{\it INTEGRAL}}
\newcommand\INTEGRAL{{\it INTEGRAL}}

\newcommand\gaia{{\it Gaia}}

\newcommand\swift{{\it Swift\/}}
\newcommand\nustar{{\it NuSTAR}}

\newcommand{\ms}{$M_{\odot}$}

\newcommand{\fluxcgs}{erg~s$^{-1}$~cm$^{-2}$}
\newcommand{\lumcgs}{erg~s$^{-1}$}

\title{Classification of IGR J20084+3221 as an Intermediate Polar using X-ray and Optical Observations}

\author[Gerber et al.]{Julian Gerber$^{1}$\thanks{E-mail: jmgerber@ucsd.edu}, Jeremy Hare$^{2,6,7}$, John A. Tomsick$^{3}$, Daniel Stern$^{4}$, Aarran W. Shaw$^{5}$\\
$^{1}$University of California, San Diego, 9500 Gilman Drive, La Jolla CA 92093, USA\\
$^{2}$NASA Goddard Space Flight Center, Greenbelt, MD 20771, USA\\
$^{3}$Space Sciences Laboratory, 7 Gauss Way, University of California, Berkeley, CA 94720-7450, USA\\
$^{4}$Jet Propulsion Laboratory, California Institute of Technology, Pasadena, CA 91109, USA\\
$^{5}$Department of Physics \& Astronomy, Butler University, 4600 Sunset Avenue, Indianapolis, IN 46208, USA\\
$^{6}$Center for Research and Exploration in Space Science and Technology, NASA/GSFC, Greenbelt, Maryland 20771, USA\\
$^{7}$The Catholic University of America, 620 Michigan Ave., N.E. Washington, DC 20064, USA
}

\begin{document}

\def\lsim{\mathrel{\lower .85ex\hbox{\rlap{$\sim$}\raise
.95ex\hbox{$<$} }}}
\def\gsim{\mathrel{\lower .80ex\hbox{\rlap{$\sim$}\raise
.90ex\hbox{$>$} }}}

\pagerange{\pageref{firstpage}--\pageref{lastpage}}

\maketitle

\label{firstpage}

\begin{abstract}
\noindent
IGR J20084+3221 is a previously unclassified Galactic source first detected by \integral. \chandra\ observations led to possible classifications of either a magnetic Cataclysmic Variable (mCV) or high mass X-ray binary (HMXB) based on the hardness of its spectrum. Here, we report follow-up observations taken by \xmm, \nustar\, and the Hale Telescope at Palomar Observatory. Based on these observations, we conclude that IGR J20084+3221 is most likely an Intermediate Polar (IP) type mCV. Timing analysis of the X-ray data found a significant peak period of $P=635.0\pm0.4$\,s, which we interpret to be the spin period of the white dwarf (WD). The X-ray spectrum is well fit to an absorbed Bremsstrahlung model with components accounting for partial covering, reflection, and a fluorescent Fe-line, all typical for an IP. The optical spectrum shows clear emission lines, consistent with emission dominated by an accretion disk. We find counterparts to the source across the optical and infrared (IR) bands, and, despite uncertainties in the distance and extinction, we estimate that the source is too faint in the IR to be an HMXB. Given the evidence pointing towards an IP classification, we fit the X-ray spectrum to a post-shock region model where we find a WD mass of $M=1.09^{+0.12}_{-0.11}$\ms, larger than the average mass for a WD in an mCV. 
\end{abstract}
\begin{keywords}
stars: white dwarfs, novae, cataclysmic variables -- accretion, accretion discs
\end{keywords}

\section{Introduction}
\label{Intro}
Cataclysmic Variables (CVs) are binary systems containing a main sequence donor star and an accreting white dwarf (WD). Of particular interest for their copious hard X-ray emission are the magnetic CVs (mCVs). Intermediate Polars (IPs) are a class of mCVs in which the magnetic field of the WD truncates the accretion disk within a critical magnetospheric radius. Within this radius, the accreted material flows along the WD magnetic field lines, ultimately falling onto the magnetic poles of the WD and emitting X-rays in the process (see e.g., \citealt{2017PASP..129f2001M} for a review). 

Hard X-ray surveys conducted by the International Gamma-Ray Astrophysics Laboratory (\INTEGRAL) have successfully expanded the known population of mCVs \citep{2022MNRAS.510.4796K}. \integral\ has proven particularly successful at detecting IPs with the highest mass WDs. IGR J14091$-$6108 and IGR J18434$-$0508, for example, have WD mass estimates approaching the Chandrasekhar limit \citep{tomsick_identifying_14091_2016, gerber_x-ray_2024}. The average mass of WDs in CVs is consistently found to be greater than that of non-accreting WDs, with $\bar M_{\mathrm{WD,~CV}}\sim0.8$\ms~and $\bar M_{\mathrm{WD,~isolated}}\sim0.5-0.7$\ms~\citep{2020MNRAS.498.3457S, Zorotovic2011A&A...536A..42Z}. The reason behind this discrepancy remains debated. Some models suggest that novae on the surface of the WD should typically expel more mass than is accreted, leading to an overall decrease in the mass of accreting WDs \citep{Yaron2005ApJ...623..398Y}. Others, however, find that WDs can experience net mass growth from accretion and novae \citep{starrfield_carbonoxygen_2020}. The lack of consensus on the evolution of these high-mass WDs necessitates their continued discovery, classification, and characterization.

Efforts to classify \integral\ sources involve follow-up soft X-ray observations in order to successfully localize the source, enabling the identification of optical and infrared (IR) counterparts. Multiwavelength photometry, coupled with X-ray spectroscopy, can then be used to classify the source \citep{2020ApJ...889...53T}. This classification schema was used to identify several \integral\ sources as IPs, including IGR J14091$-$6108, IGR J18007$-$4146, IGR J15038$-$6021, and IGR J18434$-$0508 \citep{tomsick_identifying_14091_2016, 2021ApJ...914...85H, coughenour_classifying_18007_2022, 2023MNRAS.523.4520T_15038, gerber_x-ray_2024}. Here, we present a similar observing and analysis program for IGR J20084+3221 (hereafter, \src). 

\src\ was first detected in the 14-year \integral\ catalog in the 17$-$60 keV band \citep{2017MNRAS.470..512K_integral14}. Follow up \chandra\ observations by \citet{2021ApJ...914...48T} confidently identified the counterpart CXOU J200844.1$+$321818 with position $\mathrm{R.A.}=20^\mathrm h08^\mathrm m44.14^\mathrm s, \mathrm{Dec}=32^\circ18'18.3''$. Using this more precise localization, several multiwavelength counterparts were determined, including from \gaia. Despite the known \gaia\ counterpart, the distance is poorly constrained. Therefore, a confident estimate of the absolute IR magnitude could not be determined. Given the hardness of the \chandra\ spectrum and the uncertainty on the IR magnitude, \citet{2021ApJ...914...48T} concluded that \src\ is either an mCV or high mass X-ray binary (HMXB). Here, we utilize follow-up X-ray and optical observations to classify \src\ as an IP. 

In Section \ref{Obs}, we describe the follow-up observations taken by \nustar, \xmm, and the Hale Telescope at Palomar Observatory. In Section \ref{timing}, we detail the results of the X-ray timing analysis where we determine what we interpret to be the WD spin period. In Section \ref{spectral}, we perform joint spectral analysis of the \nustar\ and \xmm\ observations to ultimately determine the mass of the WD. In Section \ref{optical}, we analyze the optical spectrum taken at Palomar as well as discuss how multiwavelength photometry supports the CV classification. Finally, in Section \ref{discussion}, we discuss how each line of evidence (timing analysis, spectral analysis, photometry) lends itself to classifying \src\ as an IP. 

\section{Observations and Data Reduction}
\label{Obs}
\src\ was observed using both \nustar\ and \xmm\ beginning on UTC 2022 May 28 and UTC 2022 May 29, respectively. An optical spectrum was taken by the Double Spectrograph on the Hale Telescope at Palomar Observatory on UT 2022 May 30. Data from each of \nustar's Focal Plane Modules (FPMA and FPMB) as well as each of \xmm's EPIC cameras (MOS1, MOS2, and pn) were reduced to obtain lightcurves and spectra. The X-ray analysis was done with HEASoft version 6.35.2. Details of the X-ray observations are summarized in Table \ref{tab:obstab}, and information on the optical observation is described in Section \ref{PalomarData}. 

\subsection{\textit{XMM-Newton}}
\label{XMMData}
The EPIC/MOS and EPIC/pn data were reduced using the \xmm\ Science Analysis Software version 22.1.0. The MOS1/2 and pn event lists were filtered using the standard filtering expressions.\footnote{``(PATTERN $<$= 12)\&\&(PI in [200:12000])\&\&\#XMMEA\_EM.'' and ``(PATTERN $<$= 4)\&\&(PI in [200:15000])\&\&(FLAG == 0),'' for MOS1/2 and pn, respectively. See~\href{https://www.cosmos.esa.int/web/xmm-newton/sas-threads\#}{https://www.cosmos.esa.int/web/xmm-newton/sas-threads\#} for reference.}
To account for the possibility of soft proton contamination, we filtered the MOS1/2 and pn event lists using the histogram method of the {\tt espfilt} tool. This filtering removes an additional $\sim 40\%$ of the observation's good time intervals. The number of counts per frame were roughly 0.02, 0.21, and 0.21 for pn, MOS1, and MOS2, respectively, which suggests that pile-up is not an issue (see e.g., Figure 5 in \citealt{2015A&A...581A.104J}). Next, we extracted source and background spectra/event files from the fully filtered MOS1/2 and pn event lists. The sources were extracted from circular regions of radii $15^{\prime\prime}$. The MOS1 background was extracted from an annular region centered on the source with an outer radius of $100^{\prime\prime}$ and an inner radius of $32.5^{\prime\prime}$. Circular background regions (each of radius $100^{\prime\prime}$) in a nearby, source-free area were used for MOS2 and pn in order to avoid the edge of the detector chip. Finally, response matrices were generated using the {\tt rmfgen} and {\tt arfgen} tools. 

\subsection{\textit{NuSTAR}}
\label{NuSTARData}
The \nustar\ observation's Level 1 event files were produced using NUSTARDAS v2.1.1 and CALDB 20220215. The high level data files were then extracted using the {\tt nuproducts} script. The input source regions were circular with radii of $40^{\prime\prime}$ for both FPMA and FPMB. The FPMB background was extracted from a $3.5^\prime \times5.0^\prime$ rectangular region on the same detector chip as the source. On FPMA, the source was within a region of stray light, so a smaller, circular background region (with a radius of $42.4^{\prime\prime}$) was used to appropriately subtract the stray light.

\subsection{Palomar}
\label{PalomarData}
\src\ was observed with the Double Spectrograph (DBSP) on the $200^{\prime \prime}$ Hale Telescope at Palomar Observatory on UT 2022 May 30. We obtained a single 700~s observation using the $1.5^{\prime \prime}$ wide slit at the parallactic angle. The night suffered from clouds and high humidity. We processed the data using standard procedures, obtaining relative flux calibration from spectrophotometric standard star observations obtained during UT 2022 May 24-28. 

\begin{table*}
\begin{minipage}{.75\linewidth}
\centering
\caption{Details of the X-ray Observations of \src}
\begin{tabular}{cccccc} \hline \hline
Observatory & ObsID & Instrument & Start Time (UTC) & End Time (UTC) & Exposure (ks) \\ \hline
\xmm & 0902370201 & MOS1 & 2022 May 29, 21:46:45 & 2022 May 30, 08:10:05 & 20.7 \\
" & " & MOS2 & " & " & 22.0 \\ 
" & " & pn & " & " & 15.9 \\
\nustar & 30801005002 & FPMA & 2022 May 28, 23:21:07 & 2022 May 29, 18:31:06 & 35.5 \\
" & " & FPMB & " & " & 35.3 \\
\hline
\label{tab:obstab}
\end{tabular}
\end{minipage}
\end{table*}

\section{Timing Analysis}
\label{timing}

For the timing analysis, we used the entire \xmm\ exposure (i.e., not removing the flaring particle background) to avoid introducing spurious peaks in the power-spectra due to gaps in the data. Prior to the timing analysis all arrival times were corrected to the solar system barycenter.
We then merged the $0.5-10$ keV source event lists for all EPIC detectors. We used the Z$_1^{2}$ test \citep{1983A&A...128..245B} to search for potential spin or orbital periodicity in the system. We searched a frequency range between 5$\times10^{-3}-0.19$ Hz (or periods between $\approx$5.2 s, corresponding to the Nyquist frequency of the MOS1/2 detector's time resolution, and 2000 s) with a uniform frequency grid spacing of $\Delta\nu=1/(10T_{\rm span})\approx2.8\times10^{-6}$ Hz, corresponding to an oversampling factor of 10. 

\begin{figure*}
\centering
\includegraphics[width=1.0\linewidth]{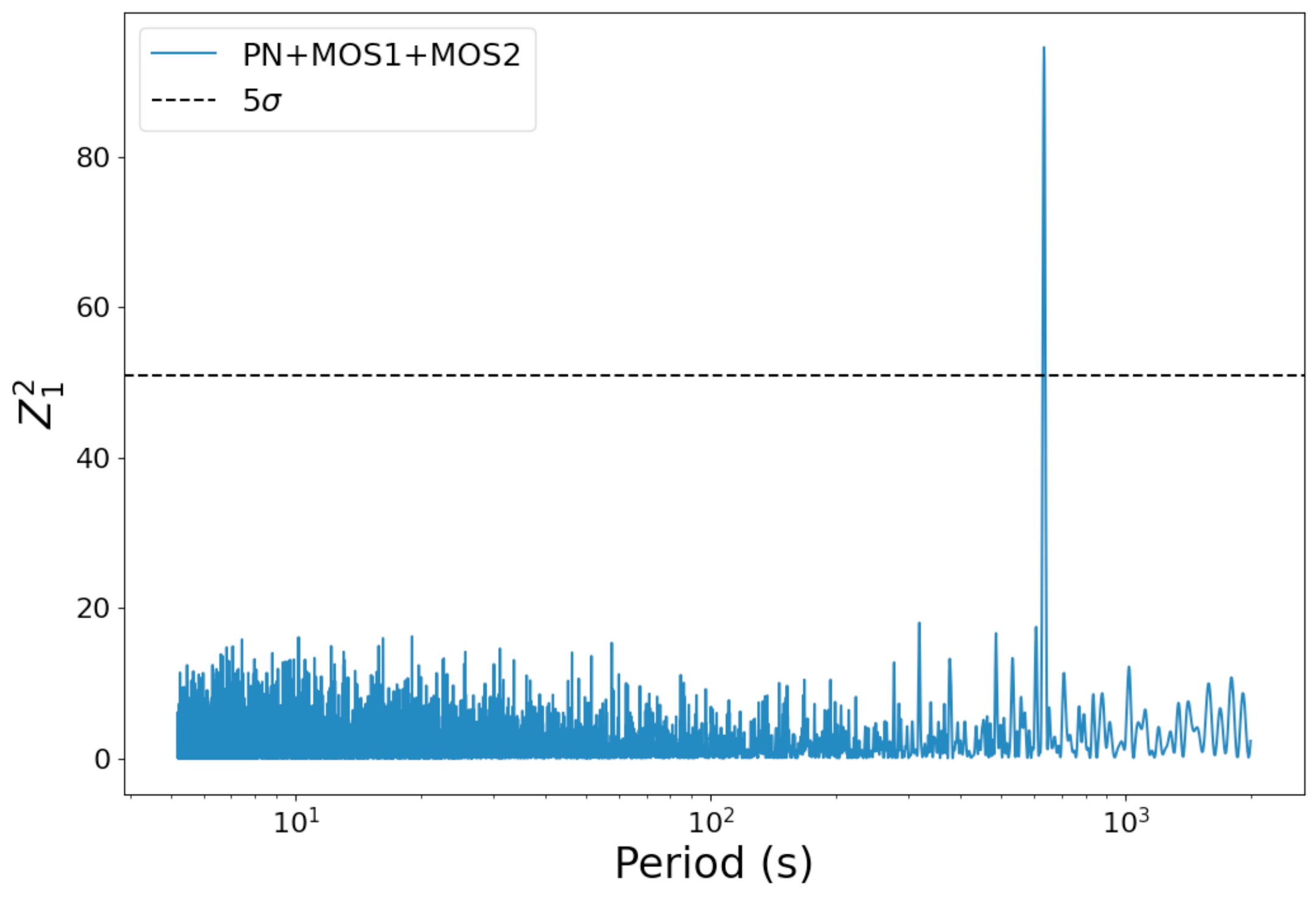}
\caption{$Z_1^2$ periodogram for the \xmm\
pn+MOS1+MOS2 data in the $0.5-10$ keV energy band. A single significant peak is found at $P=636.0\pm$0.6 s. The black dashed line shows the trials corrected $Z_1^2$ value corresponding to $5\sigma$ significance.}
\label{fig:xmmz2}
\end{figure*}

A single large peak is seen in the $Z_1^2$ periodogram at a period of $P=636.0\pm$0.6 s (or $\nu=1.572\pm0.002\times10^{-3}$ Hz), where the uncertainty in frequency was estimated using $\sigma_f=(\sqrt{3}/\pi) T^{-1}_{\rm span}(Z_{\rm 1,max}^2)^{-1/2}$ (see e.g., \citealt{2021ApJ...923..249H}). The peak's $Z_1^2$ value of 94.5 far exceeds the 5$\sigma$  trials corrected significance  and has a false alarm probability of 2$\times10^{-16}$ (see Figure \ref{fig:xmmz2}). The folded pulse profile shows a single broad peak in the $0.5-10$ keV energy range with a peak-to-peak pulse fraction of $18\pm2\%$ (see Figure \ref{fig:xmm_pulse}). We also split the event list into energies of $0.5-2$ keV and $2-10$ keV to search for energy dependence of the pulsed signal. The signal is still statistically significantly detected in the $2-10$ keV band but not in the $0.5-2$ keV band, primarily due to a lack of counts. We do not find any significant differences in the folded pulse shape between the two bands.

\begin{figure*}
\centering
\includegraphics[width=1.0\linewidth]{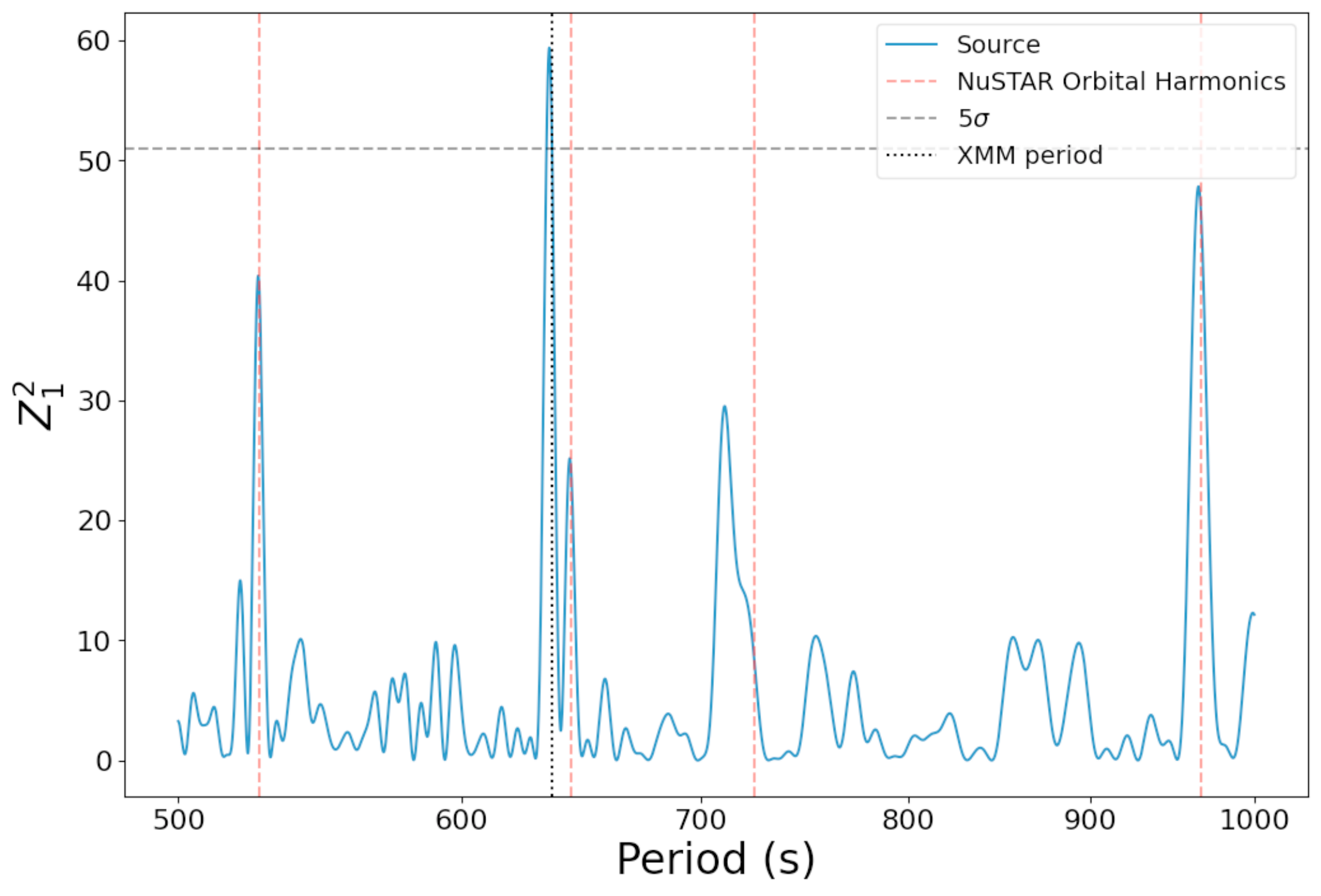}
\caption{$Z_1^2$ periodogram for the \nustar\ data in the $3-25$ keV energy band. The periodogram is zoomed in on the $500-1000$ s period range for visual clarity. A single significant peak is found at $P=635.0\pm$0.4 s in good agreement with the peak found in the \xmm\ periodogram. The black dashed line shows the trials corrected (for the full 1-1000 s period range searched) $Z_1^2$ value corresponding to $5\sigma$ significance. The black dotted line shows the period found in the \xmm\ data. The red dashed lines show where harmonics of \nustar's orbital period lie in the periodogram.}
\label{fig:nuz2}
\end{figure*}

\begin{figure}
\centering
\includegraphics[width=0.98\linewidth]{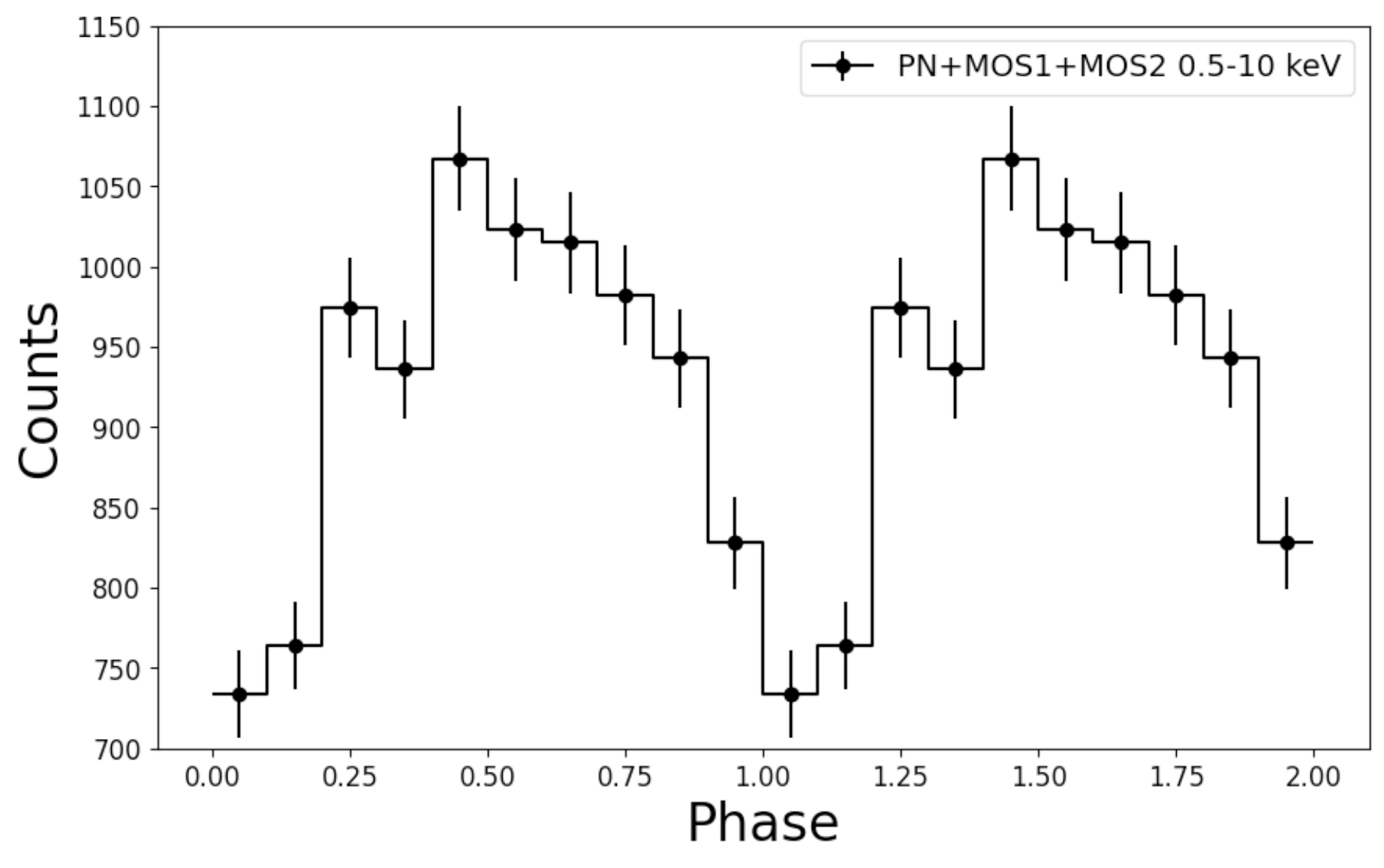}\\
\includegraphics[width=0.98\linewidth]{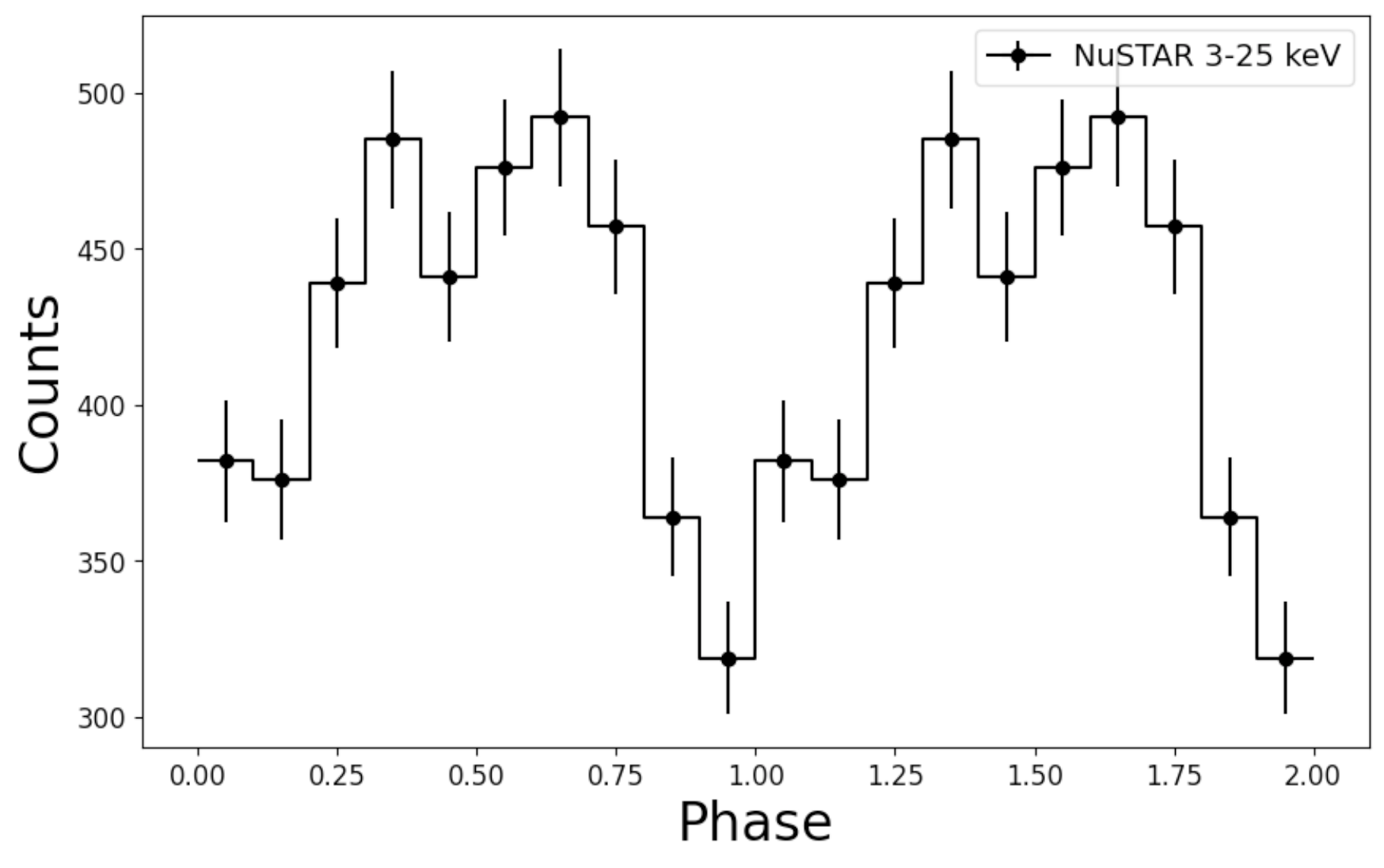}
\caption{ Top: Phase-folded \xmm\ pn+MOS1+MOS2 light curve in the $0.5-10$ keV energy band. The peak-to-peak pulsed fraction is $18\pm2\%$. Bottom: Phase-folded \nustar\ FPMA+FPMB light curve in the $3-25$ keV energy band. The peak-to-peak pulsed fraction is $22\pm3\%$}
\label{fig:xmm_pulse}
\end{figure}

We followed a similar approach to search for the period in the \nustar\ data. We corrected the event arrival times to the solar system barycenter, then filtered the event list to only contain events between $3-25$ keV. Using the $Z_1^2$ test, we searched frequencies between $1\times10^{-3}-1$ Hz with an oversampling factor of $\sim$10. The largest peak in the periodogram,  with $Z_1^2=59.4$, is found at a period of $P=635.0\pm0.4$ s (or frequency of $1.575\pm0.001\times10^{-3}$ Hz), in good agreement with the period found in the \xmm\ data. The $Z_1^2$ periodogram for the \nustar\ data is shown in Figure \ref{fig:nuz2} and is zoomed in on the period range between 500-1000 s for visualization purposes. Several other peaks are observed near the 635 s peak and are caused by harmonics of the \nustar\ orbit, which is $\approx96$ minutes. The $3-25$ keV pulse peak also shows a single broad peak and has a peak-to-peak pulse fraction of $22\pm3\%$, which is also in good agreement with the value found in the $0.5-10$ keV energy range with \xmm.

\section{X-ray Spectral Analysis}
\label{spectral}
We binned the \nustar\ and \xmm\ spectra using the \texttt{ftgrouppha} tool such that each bin had a minimum signal-to-noise ratio of 5. We fit the joint spectra above 1 keV, cutting off the \nustar\ spectra at $\sim30$ keV once the background began to dominate. To fit the spectra, we used the XSPEC package version 12.15.0d \citep{1996ASPC..101...17A}. Each model fit to the spectra included a \texttt{tbabs} component to account for absorption due to the interstellar medium (ISM). The \texttt{tbabs} model uses ISM abundances defined in \citet{2000ApJ...542..914W} and cross sections defined in \citet{1996ApJ...465..487V}. Additionally, to account for variation between the different instruments, we applied a multiplicative constant to each model. We froze the constant to 1 for the \nustar\ FPMA spectrum and left it free to vary for the others.

We began with several phenomenological models to characterize \src's spectral properties before moving on to more physically motivated models. A simple power-law model alone was unable to fit the spectra well ($\chi^2 /\nu = 766/303$). We added a Gaussian component due to the clear sign of an Fe line around $6-7$ keV in the residuals. The inclusion of a Gaussian greatly improved the fit statistic with $\chi^2/\nu = 448/300$. The best fit photon index was $\Gamma = 1.24\pm0.03$ with a Gaussian line energy of $E_{\mathrm{line}}=6.46\pm0.03$ keV (these and all subsequent confidence intervals are $1\sigma$ and are calculated using XSPEC's \texttt{error} command, unless otherwise stated). We also attempted a thermal \texttt{APEC} model designed for emission from optically thin plasmas \citep{2023AAS...24231802F}. The \texttt{APEC} model both including and excluding the Gaussian component failed to constrain the abundance and temperature.

We next moved on to more physically motivated models. Bremsstrahlung emission is often associated with IPs as a cooling mechanism for the accreted material while it falls onto the WD surface \citep{suleimanov_gk_2016}. We first fit the spectra with a thermal Bremsstrahlung model and Gaussian component which again failed to constrain the temperature and did not greatly improve the $\chi^2$ from the power-law fit ($\chi^2/\nu=427/300$). IPs are often modeled with a partial covering factor that accounts for absorption by the accretion ``curtain" formed as the material funnels along the WD's magnetic field lines \citep{2017PASP..129f2001M}. The inclusion of \texttt{pcfabs} improved the fit ($\chi^2/\nu=321/298$) and gave a best-fit temperature of $kT=35.7^{+6.7}_{-5.2}\mathrm{ ~keV}$. The coverage fraction was fit to $f=0.60\pm0.03$ and hydrogen column density fit to $N_\mathrm{H,pc}=(1.7^{+0.4}_{-0.3})\times10^{23}\mathrm{ cm}^{-2}$. 

The Fe fluorescence line at $E\sim6.4\mathrm{~keV}$ is due to neutral or only slightly ionized Fe and is a common feature in the spectra of mCVs. Physically, the 6.4 keV line in an mCV can be attributed to both the accretion column and scattering off of the surface of the WD \citep{ezuka_iron_1999}. The presence of the 6.4 keV line suggests that an additional reflection component may be necessary to model the spectrum. We therefore applied the \texttt{reflect} model from \citet{magdziarz_angle-dependent_1995} to the Bremsstrahlung continuum. When left to freely fit, the $rel_{\mathrm{refl}}$ parameter (equivalent to the fraction of solid angle subtended by the reflector) fit to unphysically high values (i.e. $rel_{\mathrm{refl}}>1$). We therefore, froze the reflection scaling factor to 1. The inclination was likewise poorly constrained extending across the parameter space from $\cos{i}=0.45 - 0.95$, so we froze the inclination to $\cos{i}=0.45$. In the \texttt{reflect} model, both the general abundance parameter and the iron abundance parameter are with respect to solar metallicities. We therefore equated the two parameters, resulting in a best-fit abundance of $Z = 0.15_{-0.12}^{+0.24}\mathrm{~Z}_\odot$ (confidence interval calculated using XSPEC's \texttt{steppar} command). The reflection component lowers the Bremsstrahlung temperature to $kT=26.1^{+7.3}_{-4.3}\mathrm{~keV}$. Despite its physical motivations, the inclusion of the \texttt{reflect} model only marginally improved the fit with $\chi^2/\nu=315/297$. 

To calculate the unabsorbed flux, we apply the \texttt{cflux} convolution model such that the combined XSPEC model is \texttt{constant*tbabs*pcfabs*cflux*(gaussian+reflect*bremss)}. We find that the best-fit $3-30$ keV flux is $7.0\pm0.2\times10^{-12}$ \fluxcgs. This is specifically the flux for the \nustar\ FPMA component of the spectrum. This is consistent with the $17-60$ keV flux of $6.8\pm0.8\times10^{-12}$ \fluxcgs~measured by \integral\ \citep{2017MNRAS.470..512K_integral14}. 

Given the spectrum's consistency with that of an mCV, we next attempted to determine the possible WD mass of \src. To do so, we replaced the Bremsstrahlung component with the post-shock region (PSR) model defined by \citet{suleimanov_gk_2016}. The PSR model (\texttt{ipolar} in XSPEC) is a grid model for emission by accreted material as it free falls from the WD's magnetospheric radius and is shocked above the surface of the WD. The input parameters of the PSR model are the WD mass, the ratio of the magnetospheric radius to the WD radius, and a normalization factor that depends on the distance to the source. In order to break the degeneracy between $M_{\mathrm{WD}}$ and $R_{\mathrm{m}}/R_{\mathrm{WD}}$, we use the $M_{\mathrm{WD}}-R_{\mathrm{WD}}$ relationship from \citet{1972ApJ...175..417N} and assume that $R_{\mathrm{m}}$ equals the co-rotation radius $R_{\mathrm{c}}=\left(\frac{GMP^2}{4\pi^2}\right)^{1/3}$ where $P=636$ s. Combining these two equations removes the ratio of radii as an input parameter. We fit \texttt{ipolar} with the partial covering and reflection models, again freezing $rel_{\mathrm{refl}}$ and $\cos{i}$ to 1.0 and 0.45, respectively. The best-fit mass is $M_{\mathrm{WD}}=1.09^{+0.12}_{-0.11} \mathrm{~M}_{\odot}$ with $\chi^2/\nu=317/297$. 

\begin{figure*}
    \centering
    \includegraphics[width=1.0\linewidth]{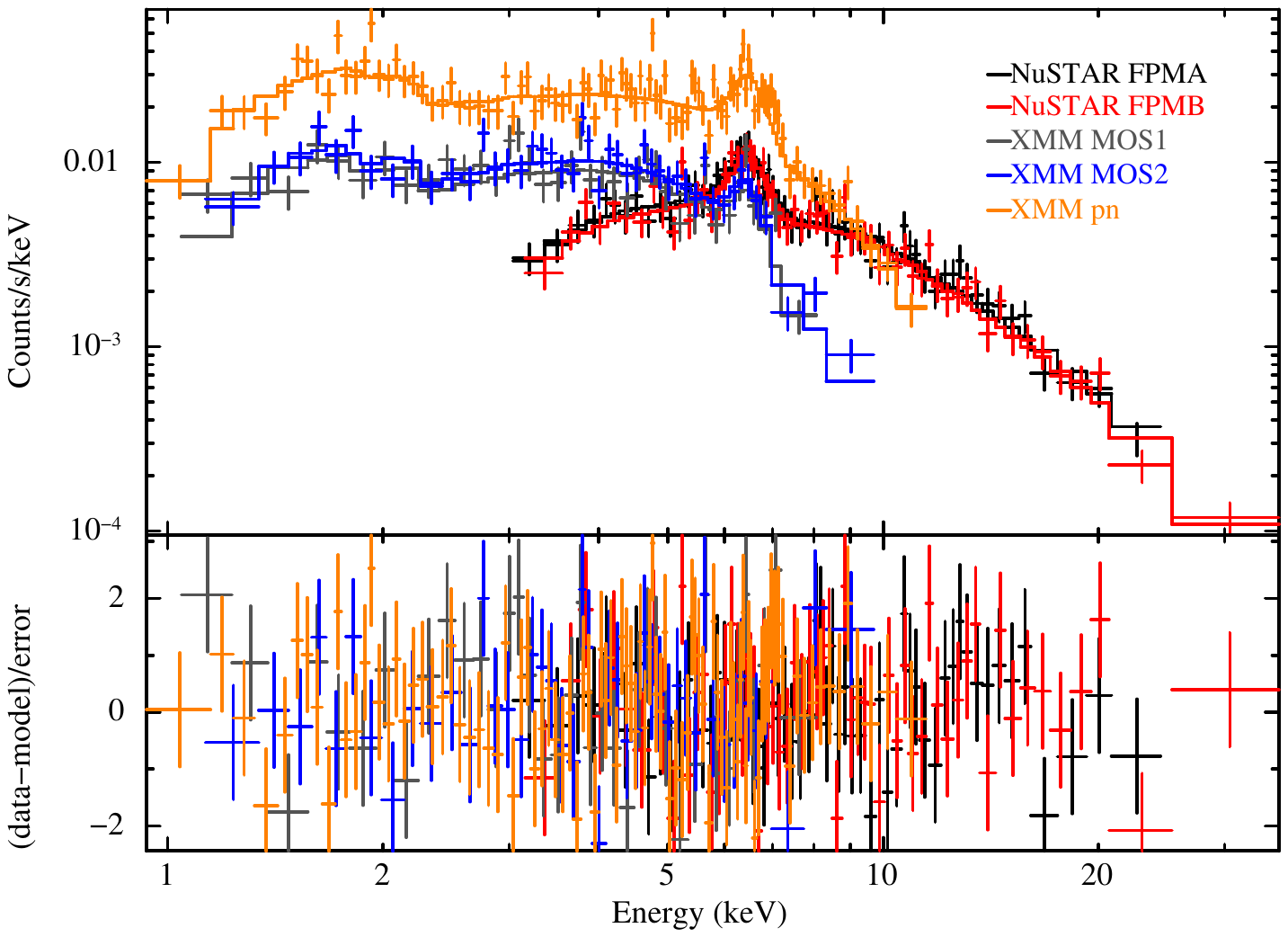}
    \caption{The joint \nustar\ and \xmm\ spectrum of \src\ fit to the post-shock region model, \texttt{ipolar}. The full model defined in XSPEC is \texttt{constant*tbabs*pcfabs*(gaussian+reflect*atable\{ipolar.fits\})}. The partial covering component, \texttt{pcfabs}, accounts for absorption by the accretion curtain while the \texttt{reflect} component models scattering off of the white dwarf surface. The resulting white dwarf mass is $M_{\mathrm{WD}}=1.09^{+0.12}_{-0.11}$ \ms with a fit statistic of $\chi^2/\nu=317/297$.}
    \label{fig:spectrum}
\end{figure*}

\section{Optical Spectrum and Multiwavelength Photometry}
\label{optical}
\cite{2021ApJ...914...48T} identified the multiwavelength counterparts to \src\ using a precise X-ray localization obtained with \chandra. The source is detected from optical to IR wavelengths and we present the broadband multiwavelength photometry in Table \ref{phottab}. The uncertain distance and extinction towards the source led \cite{2021ApJ...914...48T} to suggest that it could either be an mCV or HMXB. 

\begin{figure*}
    \centering
    \includegraphics[width=1.0\linewidth]{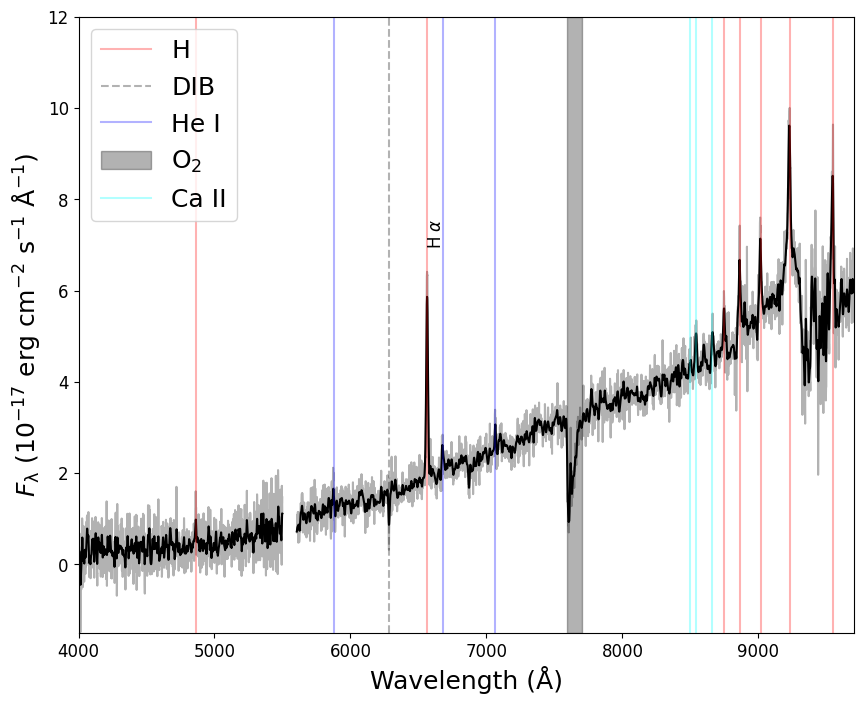}
    \caption{ Palomar optical spectrum of the counterpart of \src. The black curve is the spectrum smoothed with a 5 pixel Gaussian kernel. The fainter curve is the unsmoothed spectrum. Due to clouds at the time of the observation, the flux calibration is relative. Clear hydrogen (red), helium I (blue), and calcium II (cyan) emission lines are detected. The dashed gray line shows absorption from a diffuse interstellar band (DIB $\lambda$6281), while the gray band shows absorption from atmospheric oxygen. The gap around $\lambda$5600 is due to the dichroic that splits the two beams of the Double Spectrograph.}
    \label{fig:optspec}
\end{figure*}

We obtained an optical spectrum of this counterpart which is shown in Figure \ref{fig:optspec}. The spectrum is fairly red, consistent with the broadband photometry and relatively large inferred $N_H$ from the X-ray spectral fitting. There are clear hydrogen ($H\alpha,H\beta$, and Paschen series), He I, and Ca II triplet emission lines detected in the spectrum, but no convincing He II lines are observed. The $H\alpha$ equivalent width (EW) is $\sim36$ Å and FWHM $\sim14$ Å. There also do not appear to be any strong absorption features, except for a diffuse interstellar band (DIB) feature and atmospheric oxygen. The strong and numerous emission lines detected in the spectrum suggest that it is dominated by emission from the accretion column and/or disk. While the spectrum does appear red, there are no strong Na I D lines, which are typically associated with Galactic dust extinction.

The X-ray absorbing column density derived from the spectral fits is about $N_{\rm H}\approx1.8(2)\times10^{22}$ cm$^{-2}$ (see Table \ref{spectratable}). This corresponds to an $A_V\approx8$ or $E(B-V)\approx2.6$ for an assumed $R_V=3.1$, based on the relation of \cite{2009MNRAS.400.2050G}. Using the {\tt mwdust} package \citep{2016ApJ...818..130B} and 3D extinction map of \cite{2019ApJ...887...93G}, this corresponds to a distance of 5.2 kpc, with a range between 4-6.5 kpc based on the uncertainties in $N_{\rm H}$. However, we note that there can be large scatter in the $N_{\rm H}$ to $A_V$ and $A_V$ to distance relations based on which 3D extinction map is used, so this distance estimate should be considered with caution. In any case, the lack of significant \gaia\ parallax measurement suggests the source is at a distance of at least a few kpc. Additionally, despite the uncertainties inherent to the calculation, the distance based on extinction agrees rather well with the distance of 5.3$^{+1.5}_{-1.9}$ kpc inferred from the \gaia\ parallax \citep{2021AJ....161..147B}.  At this distance, after correcting for extinction following the conversions of $A_V$ to the \gaia\ bands from \cite{2023AJ....165..163C}, used for later comparison, the source would have an absolute \gaia\ $G$ magnitude of $M_{\rm G}\approx-0.9$ and a very blue color (i.e., \gaia\ $BP-RP\approx-0.9$). This suggests that the optical emission is dominated by the accretion disk, consistent with the emission lines seen in the optical spectrum. Furthermore, the absolute magnitude of the IR counterpart to the source is generally too faint to be consistent with a Be star at $\sim$5 kpc (see e.g., \citealt{2022RNAAS...6..163G}). Therefore, we favor the CV scenario for the nature of the source.

\section{Discussion and Conclusions}
\label{discussion}
Each line of evidence presented from the follow-up X-ray and optical observations contributes to the classification of \src\ as an IP. First, timing analysis of the X-ray data revealed a characteristic period at $P\sim635$ s. We interpret this to be the spin period of the compact object which places \src\ squarely in the typical range of WD spin periods in IPs \citep{suleimanov_hard_2019}. If we instead consider the possibility that the detected period represents the system's orbital period, then \src\ could be an ultra-compact X-ray binary (UCXB). However, an analysis of 16 UCXBs by \citet{ucxb2021MNRAS.501..548K} demonstrated that UCXBs typically show either no or weak Fe lines ($\mathrm{EW}\sim50-200$ eV). The strong Fe line evident in the X-ray spectrum of \src\ (see Figure \ref{fig:spectrum}) clearly rules out the possibility of a UCXB. 

Besides a UCXB, an interacting double WD binary would also have a very short orbital period, necessary for mass transfer to occur between the WDs. Specifically, short period binaries with stable mass transfer are expected to have orbital periods of $\lesssim20$ min \citep{Burdge2023ApJ...953L...1B}. We can infer that \src\ exhibits stable mass transfer since no outbursts have been detected and the X-ray flux does not appear variable when comparing the \nustar\ and \integral\ fluxes. However, double white dwarf binaries are typically classified as super-soft X-ray sources \citep{Maitra2024A&A...683A..21M}. This is in stark contrast to the hard X-ray spectrum shown here and in \citet{2021ApJ...914...48T}. We therefore conclude that the detected period represents the WD spin period and that the companion is likely not another WD.

The X-ray spectral analysis further contributes to the IP classification. We confirm the finding of \citet{2021ApJ...914...48T} that \src\ exhibits a hard X-ray spectrum, characteristic of an IP. In particular, we apply a thermal Bremsstrahlung model to extract a plasma temperature of $kT=26.1^{+7.3}_{-4.3}\mathrm{~keV}$ and a $3-30$ keV flux of $7.0\pm0.2\times10^{-12}$ \fluxcgs, consistent with the $17-60$ keV flux reported in the 14 year \integral\ catalog \citep{2017MNRAS.470..512K_integral14}. If we adopt a distance estimate of $4-6.5$ kpc, we find a luminosity of $(1.3-3.5)\times10^{34}$ \lumcgs. This luminosity is typical of \integral\ detected CVs, but is on the higher end of the population (see e.g., Figure 6 in \citealt{2020NewAR..9101547L}).

The X-ray analysis on its own does not definitively eliminate the possibility that \src\ is an HMXB. In some cases, neutron stars in HMXBs can have spin periods of hundreds of seconds (see e.g., \citealt{2006A&A...447.1027B}) as well as X-ray spectra with a hard continuum and strong Fe emission lines \citep{hmxb2023hxga.book..143F}. It is the optical spectroscopy and multiwavelength photometry that rules out the HMXB scenario. \src\ does not have the bright IR counterpart that would be indicative of an HMXB, and its optical spectrum appears to be dominated by an accretion disk, as opposed to a massive O or B type star. 

Based on the estimate of the source's dereddened absolute \gaia\ $G$ band magnitude and \gaia\ $BP-RP$ color, \src\ lies mostly among others with orbital periods longer than 2.6 hours \citep{2023AJ....165..163C}. 
Given that the optical spectrum is dominated by the disk, it is hard to infer much about the companion. Additionally, without a NIR/IR spectrum, it is unclear whether the emission detected by 2MASS and WISE (see Table \ref{phottab}) is from the disk or if there is some contribution from the companion star. Future NIR/IR spectra could help to place tighter constraints on the companion's spectral type.

\src\ does show some similarity to Swift J0939.7--3224, which is a CV discovered by the \swift-BAT survey \citep{2025ApJ...989..161L}. With a distance of $\approx3$ kpc \citep{2021AJ....161..147B}, Swift J0939.7--3224 was detected from optical to IR wavelengths. Follow-up optical spectroscopy of Swift J0939.7--3224 by \citet{2015AJ....150..170H} found an 8.5 hr orbital period. They additionally found relatively weak H$\alpha$ emission ($\mathrm{EW}\approx22$ Å) and no clear evidence of the companion star. These similarities suggest that \src\ may have a comparable orbital period, which could be measured by future optical radial velocity measurements.

The classification of \src\ as an IP provides additional evidence that the group of sources discovered via the \integral\ hard X-ray survey during its $>$22-year mission is a very good place to find new IPs.  As exposure time has increased throughout the mission and fainter persistent sources have been detected, it has become much more common for the Galactic sources to be IPs even though HMXBs were common in the early mission.  With the mission ending in 2025, it has now reached its maximum sensitivity, and one would expect that producing a final hard X-ray source catalog along with follow-up X-ray and optical observations will yield additional IPs.

Although an \integral\ IP population study is beyond the scope of this paper, \src\ is the fifth IGR IP that we have studied in depth with \xmm\ and \nustar, leading to X-ray measurements of the WD spin period and mass.  The WD spin period measured for \src\ ($P=635.0\pm0.4$ s) is relatively typical of IPs, and it is only slightly longer than the three slowest spinning IGR IPs that we have studied, which range from 304.4\,s to 576.3\,s.  Among our five IGR IPs, IGR J15038--6021 is the outlier with a period of 1678\,s \citep{2023MNRAS.523.4520T_15038}.  For WD masses, three of our IGR IPs are consistent with the Chandrasekhar mass, but IGR~J20084's mass of $M_{\mathrm{WD}}=1.09^{+0.12}_{-0.11} \mathrm{~M}_{\odot}$ is more similar to IGR J18007--4146 with a mass of 
$1.06\mathrm{~M}_{\odot}$ \citep{coughenour_classifying_18007_2022}.  However, it should be noted that all five IGR IPs have high masses compared to the overall IP population \citep{gerber_x-ray_2024}, which is consistent with their stronger hard X-ray emission.

Finally, while X-ray observations are critical for measuring the spin and mass, this study also shows the importance of optical spectroscopy. Given some of the observational similarities between HMXBs and IPs in the X-ray, it is the optical spectrum that has shown the presence of emission lines from an accretion disk and therefore has provided strong evidence in favor of \src\ being an IP.

\section*{Acknowledgements}
We thank Thomas Connor for providing the Palomar spectrum. JH thanks Brad Cenko and Brendan O'Connor for useful discussions regarding the analysis of optical/near-IR spectra. J.H. acknowledges
support from NASA under award number 80GSFC24M0006.  JAT acknowledges partial support from National Aeronautics and Space Administration (NASA) under \xmm\ Guest Observer grant 80NSSC22K1486. This work has made use of data from the \nustar\ mission, a project led by the California Institute of Technology, managed by the Jet Propulsion Laboratory, and funded by the National Aeronautics and Space Administration.

\section*{Data Availability}
The X-ray data used in this paper are available through NASA's HEASARC.
\label{lastpage}
\bibliography{main.bib}
\bibliographystyle{mnras.bst}
\clearpage

\begin{table*}
\caption{Spectral results for Bremsstrahlung and PSR model fits}
\begin{minipage}{\linewidth}
\begin{tabular}{cccc} \hline \hline
Parameter\footnote{The errors on the parameters are $1\sigma$ confidence intervals.} &
Units &
Bremsstrahlung Model\footnote{The full XSPEC model is {\tt{constant*tbabs*pcfabs*(gaussian+reflect*bremss)}}.} &
PSR Model\footnote{The full XSPEC model is {\tt{constant*tbabs*pcfabs*(gaussian+reflect*atable\{ipolar.fits\})}}.} \\ \hline
$N_{\rm H}$ & $10^{22}$ cm$^{-2}$ & $1.76\pm0.18$  & $1.86\pm0.17$ \\
$N_{\rm H,pc}$ & $10^{22}$ cm$^{-2}$  & $12.01^{+4.0}_{-3.0}$ & $14.43^{+4.5}_{-3.1}$\\
pc fraction & --- & $0.52\pm0.07$ & $0.55\pm0.06$ \\
\hline
$E_{\rm line}$ & keV & $6.48\pm0.03$ & $6.48\pm0.03$ \\
$\sigma_{\rm line}$ & keV  & $0.29\pm0.04$ & $0.28\pm0.04$\\
$N_{\rm line}$ & ph cm$^{-2}$ s$^{-1}$  & $(2.7\pm0.3)\times10^{-5}$ & $(2.7\pm0.3)\times10^{-5}$ \\
$EW_{\rm line}$\footnote{The error is estimated by assuming that the equivalent width and line normalization have equal fractional errors. The quoted values apply for the \nustar\ FPMA spectrum, though the other spectra's equivalent widths vary only by a few eV.} & eV & $694 \pm 77$ & $680 \pm 76$\\
\hline
$rel_{\rm refl}$ & ---  & 1.0\footnote{\label{frozen}Frozen.} &  1.0$^e$\\
$Z$ & --- &  $0.15_{-0.12}^{+0.24}$\footnote{Confidence interval determined using XSPEC's \texttt{steppar} command.} &   $0.11_{-0.08}^{+0.16}$$^f$ \\
$Z_{\rm Fe}$\footnote{Tied to parameter $Z$.} & --- & 0.15 & 0.11 \\
$\cos{i}$ & --- & 0.45$^e$ & 0.45$^e$  \\ 
\hline
$kT$ & keV & $26.1^{+7.3}_{-4.3}$ & ---\\
$N_{\rm bremss}$ & --- & $6.2^{+1.0}_{-0.9}\times10^{-4}$ & --- \\
\hline
$M_{\rm WD}$ & \ms & --- & $1.09^{+0.12}_{-0.11}$ \\
$R_{\rm m}$\footnote{Linked to $M_{\rm WD}$ via \citet{1972ApJ...175..417N} and assumption that $R_{\rm m}$ equals the corotation radius.} & $R_{\rm WD}$ & --- & 24.1 \\
$N_{\rm PSR}$ & --- & --- & $6.8^{+6}_{-4}\times10^{-29}$ \\
\hline
C$_{\rm FPMA}$ & --- & 1.0$^e$ &  1.0$^e$ \\
C$_{\rm FPMB}$ & --- & $0.98\pm0.03$ & $0.98\pm0.03$ \\ 
C$_{\rm MOS1}$ & --- & $0.71\pm0.03$ &  $0.71\pm0.03$ \\
C$_{\rm MOS2}$ & --- & $0.75\pm0.03$ & $0.74\pm0.03$ \\
C$_{\rm pn}$ & --- & $0.74\pm0.03$ & $0.74\pm0.03$ \\
\hline
$\chi^2/\nu$ & --- & $315/297$ & $317/297$\\
\hline
\end{tabular}
\end{minipage}
\label{spectratable}
\end{table*}

\begin{table*}
\caption{Broadband Photometry}
\begin{minipage}{1\linewidth}
\centering
\begin{adjustbox}{max width=\linewidth}
\begin{tabular}{ccccccccccccc} \hline \hline
& \gaia\ DR3\footnote{\cite{2016A&A...595A...1G,2023A&A...674A...1G}} & & & & PanSTARRS\footnote{\cite{2020ApJS..251....6M,2025yCat.2389....0M}} & & & UKIDSS\footnote{\cite{2008MNRAS.391..136L,2012yCat.2316....0U}} & & & AllWISE\footnote{\cite{allwise2014yCat.2328....0C}} &  \\ \hline
$G$\footnote{The \gaia, UKIDSS, and AllWISE surveys use Vega magnitudes while the PanSTARRS survey uses AB magnitudes. } & $BP$ & $RP$ & $g$ & $r$ & $i$ & $z$ & $y$ & $J$ & $H$ & $K$ & W1 & W2   \\ \hline
19.441(6) & 20.89(8) & 18.22(2) & 21.43(6) & 19.628(4) & 18.69(4) & 18.2(1) & 17.65(2) & 16.037(7) & 15.306(6) & 14.89(1) &14.60(7) & 14.57(8) \\
\hline
\end{tabular}
\end{adjustbox}
\end{minipage}
\label{phottab}
\end{table*}

\end{document}